\pdfoutput=1
\documentclass[aps,pra,reprint]{revtex4-2}

\usepackage[justification=justified,singlelinecheck=false]{caption}
\usepackage{float}
\usepackage{amsmath,amssymb,amsfonts}
\usepackage{graphicx}
\usepackage{overpic}
\usepackage{bm}
\usepackage[left]{lineno}
\makeatletter
\renewcommand{\paragraph}[1]{%
  \par\vspace{0.6em}%
  \noindent\textit{#1 ---}
}
\makeatother

\usepackage{color}

\usepackage{braket}
\usepackage{esint}
\usepackage{dutchcal}
\usepackage[utf8]{inputenc}

\usepackage{xcolor}

 \usepackage[unicode=true,pdfusetitle,
bookmarks=true,bookmarksnumbered=false,bookmarksopen=false,breaklinks=true,pdfborder={0 0 1},backref=false,colorlinks=true]
  {hyperref}
 \hypersetup{
 linkcolor=red,urlcolor=blue,citecolor=red,anchorcolor=blue}

\makeatletter
\@ifundefined{textcolor}{}
{%
 \definecolor{BLACK}{gray}{0}
 \definecolor{WHITE}{gray}{1}
 \definecolor{RED}{rgb}{1,0,0}
 \definecolor{GREEN}{rgb}{0,1,0}
 \definecolor{BLUE}{rgb}{0,0,1}
 \definecolor{CYAN}{cmyk}{1,0,0,0}
 \definecolor{MAGENTA}{cmyk}{0,1,0,0}
 \definecolor{YELLOW}{cmyk}{0,0,1,0}
}

\usepackage{xcolor}
\usepackage{ifthen}
\usepackage{marginnote}

\newboolean{showedits}
\setboolean{showedits}{true}

\usepackage[normalem]{ulem}

\newif\iferbedit
\erbedittrue   
\erbeditfalse 

\iferbedit
  
  \newcommand{\erbdel}[1]{\textcolor{red}{\sout{#1}}}
  
\else
  
  \newcommand{\erbdel}[1]{}
  
\fi

\usepackage{times}
\usepackage{psfrag}
\usepackage{epsf}
\usepackage{epsfig}
\usepackage{latexsym}
\usepackage{amsfonts}
\usepackage{graphicx}
\usepackage{subcaption}
\usepackage{color}
\usepackage{epsf}
\usepackage{latexsym}
 \usepackage{ifpdf}

\makeatother

\begin{document}

\title{Transient Synchronization and Entanglement through Shared Dissipation in Circuit QED}

\author{Areeda Ayoub} 
\affiliation{Department of Physics,
University of Houston, Houston TX, USA}

\author{Eric R. Bittner}
\email{ebittner@central.uh.edu}
\affiliation{Department of Physics,
University of Houston, Houston TX, USA}

 \date{\today}

\begin{abstract}
We propose a superconducting circuit in which two resonators
couple through DC-biased SQUIDs to a common lossy auxiliary
mode. Adiabatic elimination yields collective damping and
correlated noise with an overlap $\xi$ whose magnitude is tuned
by the flux biases and whose sign follows the relative sign of
the couplings. The full three-mode dynamics confirm this reduced
description. For detuned resonators, we derive the
exceptional-point condition, above which the mean amplitudes
lock to a common frequency. For resonant modes, unequal
collective decay produces transient synchronization, antiphase
or in phase depending on the relative sign of the couplings,
with an analytical synchronization time set by the collective
decay splitting and the initial mean amplitudes. For initially
separable squeezed states, the same relaxation converts local
squeezing into transient entanglement, quantified by
logarithmic negativity. We obtain closed-form expressions for
the squeezing threshold above which entanglement appears and
for the time at which it disappears. Thermal noise suppresses
entanglement without changing the mean-amplitude
synchronization time at fixed parameters. For realistic circuit
parameters, these effects occur within a few hundred
nanoseconds and are accessible through output-field
measurements. A single tunable shared channel thus produces an
exceptional point, transient synchronization, and transient
entanglement, and provides a controllable source of correlated
noise for superconducting circuits.
\end{abstract}

\maketitle
\section{Introduction}
\label{sec:introduction}

In 1665, Christiaan Huygens observed antiphase
synchronization between two pendulum clocks coupled
through a common support~\cite{Bennett2002}.
Modern experiments with clocks and metronomes demonstrate
how the coupling structure selects synchronized
motion~\cite{Pantaleone2002,PenaRamirez2016}.
In open quantum systems, a common reservoir provides a
related mechanism: unequal damping of collective modes
allows a slowly decaying contribution to dominate,
establishing a fixed phase relation between
oscillators~\cite{Giorgi2012,Manzano2013}.

Reservoir engineering uses dissipation for quantum-state
preparation~\cite{Poyatos1996,Verstraete2009}, including
entanglement stabilization in superconducting
circuits~\cite{Shankar2013}.
Bittner and Tyagi showed that correlated dissipative
environments can select long-lived phase-synchronized
motion through the symmetry of the
dissipator~\cite{BittnerTyagi2025}.
A circuit realization must connect these effective
correlations to tunable parameters and establish how
private losses and thermal fluctuations affect both
collective motion and quantum correlations.

A strongly damped auxiliary mode provides a shared
reservoir and mediates dissipative
interactions~\cite{MetelmannClerk2015,Yu2019}.
Such interactions connect reservoir engineering to
non-Hermitian spectral physics, including level
attraction and exceptional points, where eigenvalues
and eigenvectors coalesce~\cite{Harder2018,MiriAlu2019,Ashida2020}.
Exceptional points have been observed in superconducting
circuits, both in tunable resonators~\cite{Partanen2019}
and in a dissipative qubit~\cite{Naghiloo2019}.
For balanced losses and opposite detunings, dissipative
coupling realizes anti-parity--time symmetry~\cite{Peng2016}.
Exceptional points and synchronization are nevertheless
distinct: selection of a slowly decaying collective mode
does not require spectral coalescence.

Here, we propose a superconducting circuit that supports
exceptional points, transient synchronization, and
transient entanglement through tunable shared dissipation.
Two system resonators couple to a lossy auxiliary resonator
through DC-biased SQUIDs, while independent transmission
lines provide private loss channels and readout ports. The
design uses flux-tunable resonator couplings~\cite{Peropadre2013}
and established circuit-QED elements~\cite{Cattaneo2021,BlaisRMP2021}.
All couplings are static, with no direct coherent coupling
between the system resonators. Eliminating the auxiliary
mode yields collective damping and correlated input noise.
We relate the normalized cross-decay coefficient $\xi$
to the circuit couplings and decay rates: its magnitude is set
by the flux biases, while its sign follows the relative sign of
the two couplings and determines which collective mode is
protected from the shared channel. We check the
reduced description against the full three-mode dynamics
throughout.

For detuned resonators, we derive the exceptional-point
condition at fixed private loss. Above the exceptional point,
the mean amplitudes lock to a common frequency with a fixed
relative phase. Comparison with the full three-mode spectrum
shows that the finite response time of the auxiliary mode
shifts the exceptional point by an amount that scales as
$1/\kappa_b$. For equal system frequencies, we obtain the time
required to reach a prescribed antiphase tolerance; for
couplings of opposite sign the same analysis gives in-phase
alignment. This time is set by the collective decay splitting
and the initial mean amplitudes, and numerical integration
confirms the analytical prediction.

For initially separable squeezed states, we solve the
reduced covariance dynamics and quantify entanglement using
the logarithmic negativity~\cite{Simon2000,VidalWerner2002,Weedbrook2012}.
Unequal collective relaxation converts local squeezing into
transient intermode entanglement. We derive closed-form
expressions for the squeezing threshold above which
entanglement appears and for the time at which it disappears,
and we examine the effects of shared coupling, thermal
occupation, and the amount of initial squeezing. Entanglement
and antiphase alignment can coexist, but they remain distinct:
synchronization concerns decaying first moments, whereas
entanglement concerns centered quantum fluctuations.

For realistic circuit parameters, these effects occur within
a few hundred nanoseconds and persist at thermal occupations
an order of magnitude higher than those of typical devices. We
describe how the squeezed initial states can be prepared with
a parametric pump, and how both the synchronization criterion
and the covariance matrix are obtained from output-field
quadrature measurements, including the surviving antiphase
signal at the synchronization threshold. Because $|\xi|$ can be
calibrated from two ring-down rates, the circuit also provides a
controllable source of correlated noise for studying correlated
errors in superconducting processors~\cite{Wilen2021,McEwen2022}.

Section~\ref{sec:model} presents the circuit, the reduced
dynamics, and the physical parameters.
Section~\ref{sec:results} discusses the exceptional-point
spectrum, synchronization, and entanglement.
Section~\ref{sec:experimental} addresses state preparation,
readout, and imperfections, and Section~\ref{sec:conclusion}
summarizes the findings and discusses possible applications.
\section{Model}
\label{sec:model}

\subsection{Circuit model}
\label{sec:circuit_model}

\begin{figure}[t]
\centering
\includegraphics[width=\linewidth]{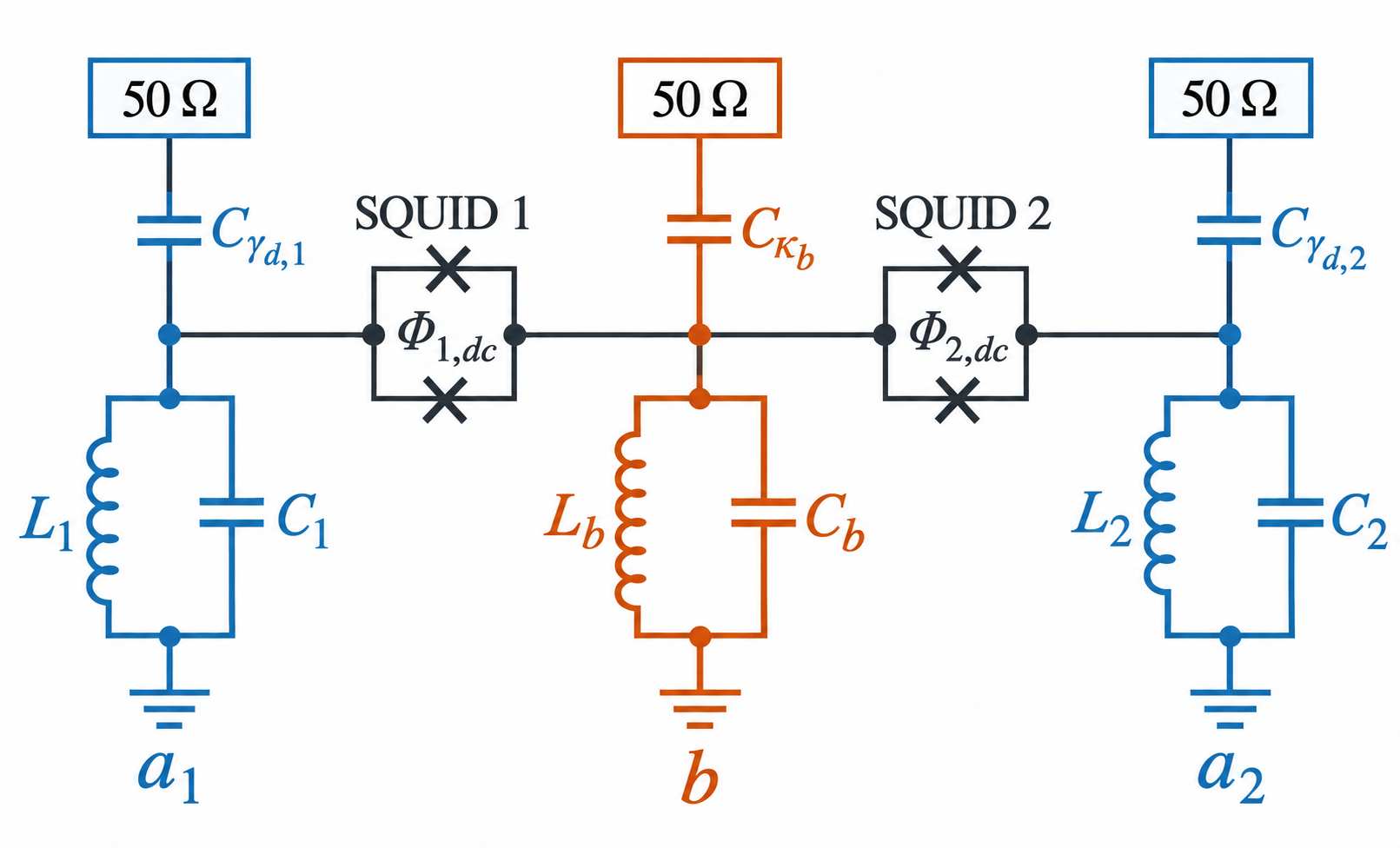}
\caption{
Circuit realizing a shared dissipation channel.
The system resonators $a_1$ and $a_2$ (blue) couple
to the lossy auxiliary resonator $b$ (orange) through
two DC-biased SQUIDs, whose fluxes $\Phi_{1,\mathrm{dc}}$
and $\Phi_{2,\mathrm{dc}}$ set $g_1$ and $g_2$.
The capacitors $C_{\gamma_{d,1}}$, $C_{\gamma_{d,2}}$,
and $C_{\kappa_b}$ couple the modes to independent
$50\,\Omega$ lines, producing the energy-decay
rates $\gamma_{d,1}$, $\gamma_{d,2}$, and $\kappa_b$.
There is no direct coupling between the system resonators.
}
\label{fig:circuit_model}
\end{figure}

We consider two superconducting resonators, $a_1$ and $a_2$,
coupled to a common lossy auxiliary resonator $b$, as shown
in Fig.~\ref{fig:circuit_model}. The system resonators are
represented by the parallel LC circuits $(L_1,C_1)$ and
$(L_2,C_2)$, and the auxiliary resonator by $(L_b,C_b)$.
Each system resonator is coupled to $b$ through a DC-biased
SQUID, whose flux bias $\Phi_{j,\mathrm{dc}}$ sets the coupling
strength $g_j$~\cite{Peropadre2013}. There is no direct coupling
between $a_1$ and $a_2$. Each resonator is also coupled through
a capacitor to an independent $50\,\Omega$ transmission line,
which produces private energy decay of the system modes and
strong damping of the auxiliary mode~\cite{Krantz2019,Cattaneo2021}.
When this damping is large, $b$ acts as a common reservoir for
$a_1$ and $a_2$.

We work in the small-phase, linear regime of the SQUIDs and
employ the rotating-wave approximation. Since the couplings are
static, $g_1$ and $g_2$ are real. Their magnitudes are set by the
flux biases, while their signs are fixed by the circuit geometry,
i.e., by the polarity with which each coupler is connected to its
resonator. We keep the signs general in Sec.~\ref{sec:effective_dynamics},
since the relative sign of $g_1$ and $g_2$ determines which collective
mode is protected from the shared channel. The resonance frequencies
include the local reactive shifts from the couplers and ports.

We use dimensionless units with the private energy-decay rate
$\gamma_d$ as the reference scale. Time is measured in units of
$\gamma_d^{-1}$, so that $\tau=\gamma_d t$; frequencies, coupling
strengths, and decay rates are measured in units of $\gamma_d$,
and energies in units of $\hbar\gamma_d$. We retain the same
symbols for the scaled parameters, and an overdot denotes
$d/d\tau$. Input fields are normalized with respect to $\tau$,
while mode operators and quadrature variances are unchanged.
Physical values are given in Sec.~\ref{sec:parameters}.

In a frame rotating at the auxiliary frequency $\omega_b$,
the Hamiltonian is
\begin{equation}
\mathcal H=\sum_{j=1,2}\left[\Delta_ja_j^\dagger a_j
+g_j\left(a_j^\dagger b+b^\dagger a_j\right)\right],
\label{eq:rotating_hamiltonian}
\end{equation}
where $\Delta_j=\omega_j-\omega_b$. The Markov quantum
Langevin equations are
\begin{align}
\dot a_1
&=-\left(\frac{\gamma_{d,1}}2+i\Delta_1\right)a_1
-ig_1b+\sqrt{\gamma_{d,1}}\,a_{1,\mathrm{in}},
\label{eq:qle_a1}\\
\dot a_2
&=-\left(\frac{\gamma_{d,2}}2+i\Delta_2\right)a_2
-ig_2b+\sqrt{\gamma_{d,2}}\,a_{2,\mathrm{in}},
\label{eq:qle_a2}\\
\dot b
&=-\frac{\kappa_b}2b-i(g_1a_1+g_2a_2)
+\sqrt{\kappa_b}\,b_{\mathrm{in}}.
\label{eq:qle_b}
\end{align}
The inputs are independent, zero-mean thermal fields. Writing
$f_1=a_{1,\mathrm{in}}$, $f_2=a_{2,\mathrm{in}}$, and
$f_b=b_{\mathrm{in}}$, their nonvanishing second moments are
\begin{align}
\langle f_\mu(\tau)f_\nu^\dagger(\tau')\rangle
&=\delta_{\mu\nu}(\bar n_\mu+1)\delta(\tau-\tau'),
\label{eq:input_antinormal}\\
\langle f_\mu^\dagger(\tau)f_\nu(\tau')\rangle
&=\delta_{\mu\nu}\bar n_\mu\delta(\tau-\tau'),
\label{eq:input_normal}
\end{align}
where $\mu,\nu\in\{1,2,b\}$. The occupations are taken
constant over the relevant bandwidth, and all anomalous
input correlations vanish.

The private lines also serve as readout ports. With the sign
convention of Eqs.~\eqref{eq:qle_a1} and~\eqref{eq:qle_a2},
the field leaving resonator $j$ is
\begin{equation}
a_{j,\mathrm{out}}=\sqrt{\gamma_{d,j}}\,a_j-a_{j,\mathrm{in}}.
\label{eq:input_output}
\end{equation}
Since the inputs have zero mean,
$\langle a_{j,\mathrm{out}}\rangle=\sqrt{\gamma_{d,j}}\,\langle a_j\rangle$,
so averaged output traces give the mean amplitudes of both
resonators directly. The fluctuations of $a_{j,\mathrm{out}}$
carry the intracavity covariances. The quantities computed in
Sec.~\ref{sec:results} are therefore accessible through the
private ports; the measurement is discussed in
Sec.~\ref{sec:experimental}.

\subsection{Effective dissipative coupling}
\label{sec:effective_dynamics}

We eliminate the auxiliary mode in the regime
\begin{equation}
\frac{\kappa_b}{2}\gg
\max\left\{\sqrt{g_1^2+g_2^2},
\frac{\gamma_{d,1}}2,\frac{\gamma_{d,2}}2,
|\Delta_1|,|\Delta_2|\right\}.
\label{eq:adiabatic_condition}
\end{equation}
Setting $\dot b\simeq0$ in Eq.~\eqref{eq:qle_b} gives
\begin{equation}
b\simeq-\frac{2i}{\kappa_b}(g_1a_1+g_2a_2)
+\frac{2}{\sqrt{\kappa_b}}\,b_{\mathrm{in}}.
\label{eq:eliminated_b}
\end{equation}
Keeping the detunings in the response of $b$ replaces $2/\kappa_b$
by $1/(\kappa_b/2-i\Delta_j)$. The resulting corrections, of relative
order $2|\Delta_j|/\kappa_b$, introduce frequency shifts and a coherent
exchange between $a_1$ and $a_2$, and are neglected here.
Substitution gives the reduced equations
\begin{align}
\dot a_1
&=-\left(\frac{\Gamma_1}{2}+i\Delta_1\right)a_1
-\frac{\Gamma_{12}}2a_2+F_1,
\label{eq:reduced_qle_a1}\\
\dot a_2
&=-\left(\frac{\Gamma_2}{2}+i\Delta_2\right)a_2
-\frac{\Gamma_{12}}2a_1+F_2,
\label{eq:reduced_qle_a2}
\end{align}
where
\begin{align}
\gamma_{s,j}&=\frac{4g_j^2}{\kappa_b},
&\Gamma_j&=\gamma_{d,j}+\gamma_{s,j},
\label{eq:effective_local_rates}\\
\Gamma_{12}&=\frac{4g_1g_2}{\kappa_b}
=\sigma\sqrt{\gamma_{s,1}\gamma_{s,2}},
&\sigma&=\operatorname{sgn}(g_1g_2),
\label{eq:cross_damping}
\end{align}
and the effective noise forces are
\begin{equation}
F_j=\sqrt{\gamma_{d,j}}\,a_{j,\mathrm{in}}
-\frac{2ig_j}{\sqrt{\kappa_b}}\,b_{\mathrm{in}}.
\label{eq:effective_noise}
\end{equation}
The local rates $\gamma_{s,j}$ depend only on $g_j^2$, whereas the
cross-damping $\Gamma_{12}$ and the shared part of the noise carry the
relative sign of the couplings. These equations describe private damping
of each mode together with a collective dissipator with jump operator
$(2/\sqrt{\kappa_b})(g_1a_1+g_2a_2)$. The cross-damping
and the correlated fluctuations thus originate from the same channel
and must be retained together.

We characterize the shared channel by
\begin{equation}
\xi=\frac{\Gamma_{12}}{\sqrt{\Gamma_1\Gamma_2}}
=\sigma\sqrt{\frac{\gamma_{s,1}\gamma_{s,2}}
{(\gamma_{d,1}+\gamma_{s,1})(\gamma_{d,2}+\gamma_{s,2})}}.
\label{eq:xi}
\end{equation}
For positive private decay rates and finite couplings,
$|\xi|<1$: the magnitude of $\xi$ is set by the flux biases and its
sign by the relative sign of $g_1$ and $g_2$. The normalized effective
inputs $\eta_j=F_j/\sqrt{\Gamma_j}$ satisfy
\begin{equation}
[\eta_j(\tau),\eta_k^\dagger(\tau')]
=M_{jk}\delta(\tau-\tau'),
\qquad
M=\begin{pmatrix}1&\xi\\\xi&1\end{pmatrix},
\label{eq:noise_overlap}
\end{equation}
and their cross-correlations are
\begin{align}
\langle\eta_1(\tau)\eta_2^\dagger(\tau')\rangle
&=\xi(\bar n_b+1)\delta(\tau-\tau'),
\label{eq:cross_antinormal}\\
\langle\eta_1^\dagger(\tau)\eta_2(\tau')\rangle
&=\xi\bar n_b\delta(\tau-\tau').
\label{eq:cross_normal}
\end{align}
The local effective occupations are
\begin{equation}
\bar n_{j,\mathrm{eff}}=
\frac{\gamma_{d,j}\bar n_j+\gamma_{s,j}\bar n_b}{\Gamma_j},
\label{eq:effective_occupation}
\end{equation}
so that
\begin{align}
\langle\eta_j^\dagger(\tau)\eta_j(\tau')\rangle
&=\bar n_{j,\mathrm{eff}}\delta(\tau-\tau'),
\label{eq:local_normal}\\
\langle\eta_j(\tau)\eta_j^\dagger(\tau')\rangle
&=(\bar n_{j,\mathrm{eff}}+1)\delta(\tau-\tau').
\label{eq:local_antinormal}
\end{align}
Thus $\xi$ measures the overlap of the effective input
channels. For unequal reservoir occupations, it need
not equal the cross-correlation normalized by the local
thermal noise powers.

The circuit realizes the correlated dissipation of
Ref.~\cite{BittnerTyagi2025} without a direct coherent coupling
between the oscillators. The auxiliary mode couples only to the
bright combination $g_1a_1+g_2a_2$, while the orthogonal combination
\begin{equation}
a_{\mathrm D}=\frac{g_2a_1-g_1a_2}{\sqrt{g_1^2+g_2^2}}
\label{eq:dark_mode}
\end{equation}
is decoupled from the shared channel and is damped only by the
private losses. For $|g_1|=|g_2|$, this dark mode is the antisymmetric
combination $a_-=(a_1-a_2)/\sqrt2$ when the couplings have equal signs
($\xi>0$), and the symmetric combination $a_+=(a_1+a_2)/\sqrt2$ when
they have opposite signs ($\xi<0$). Because $a_1$ and $a_2$ interact
only through $b$, the two cases are related by the local phase rotation
$a_2\to-a_2$, which reverses the sign of $g_2$ and of $\xi$. This
transformation leaves the decay rates, the eigenvalue spectrum, and the
entanglement unchanged, and only shifts the relative phase of the
long-lived collective motion from $\pi$ to $0$.

Hereafter we consider symmetric private decay and
couplings of equal sign, $\gamma_{d,1}=\gamma_{d,2}=1$ and
$g_1=g_2=g_c>0$, so that $0\le\xi<1$; the results for
$g_1=-g_2$ follow from $a_2\to-a_2$. Then $\Gamma_1=\Gamma_2=\Gamma$ and
$\Gamma_{12}=\gamma_s=\xi\Gamma$. At fixed private
decay, the coupling sweep is parameterized by
\begin{equation}
\gamma_s(\xi)=\frac{\xi}{1-\xi},
\qquad
\Gamma(\xi)=\frac{1}{1-\xi},
\label{eq:xi_decay_rates}
\end{equation}
and
\begin{equation}
g_c(\xi)=\frac12\sqrt{\frac{\kappa_b\xi}{1-\xi}}.
\label{eq:xi_coupling}
\end{equation}
In this configuration the coupling condition in
Eq.~\eqref{eq:adiabatic_condition} reads
$\sqrt2\,g_c/(\kappa_b/2)\ll1$, or equivalently
$\xi/(1-\xi)\ll\kappa_b/2$, so larger values of $\xi$
require a larger $\kappa_b$.

For $\boldsymbol a=(a_1,a_2)^T$, the reduced dynamics take
the form
\begin{equation}
\dot{\boldsymbol a}=W\boldsymbol a+\boldsymbol F,
\label{eq:matrix_qle}
\end{equation}
where $\boldsymbol F=(F_1,F_2)^T$ and
\begin{equation}
W=-\frac12\begin{pmatrix}
\Gamma+2i\Delta_1&\gamma_s\\
\gamma_s&\Gamma+2i\Delta_2
\end{pmatrix}.
\label{eq:drift_matrix}
\end{equation}
The creation operators obey the complex-conjugate equations.
Both the local damping and the cross-damping vary with
$\xi$. Reservoir occupations enter the noise correlations
but not $W$; consequently, its eigenvalues and the mean
amplitudes are independent of these occupations for fixed
rates and initial first moments.

Two detuning configurations are used below. Opposite
detunings, $\Delta_1=\delta/2$ and $\Delta_2=-\delta/2$,
with $\delta=\omega_1-\omega_2$, are used to investigate
exceptional points. Resonant modes, $\Delta_1=\Delta_2=0$,
are used for synchronization and entanglement; a common
detuning would only add a common rotation in the reduced
model while increasing the neglected corrections. Varying
$\xi$ at fixed detunings and private decay assumes that the
frequency shifts accompanying changes of the SQUID flux
biases are compensated.

\subsection{Physical parameters}
\label{sec:parameters}

To connect the scaled parameters to a device, we take
resonance frequencies near $5$~GHz, a private decay rate
$\gamma_d/2\pi=0.5$~MHz, and an auxiliary decay rate
$\kappa_b/2\pi=100$~MHz, so that $\kappa_b=200$ in scaled
units. The time unit is then $1/\gamma_d\simeq0.32~\mu$s. For
$\xi=0.8$, Eq.~\eqref{eq:xi_coupling} gives
$g_c/2\pi\simeq7.1$~MHz and $\sqrt2\,g_c/(\kappa_b/2)\simeq0.20$;
for $\xi=0.9$, $g_c/2\pi\simeq10.6$~MHz and
$\sqrt2\,g_c/(\kappa_b/2)\simeq0.30$. These couplings are
well within the range of flux-tunable couplers between
superconducting resonators~\cite{Peropadre2013}. All rates
remain small compared with the resonance frequencies,
consistent with the rotating-wave and Markov approximations.
The private decay is dominated by the engineered port loss:
an internal quality factor near $10^6$ contributes an internal
decay rate of $2\pi\times5$~kHz, about $1\%$ of $\gamma_d$.
\section{Results}
\label{sec:results}

\subsection{Exceptional point}
\label{sec:spectral_properties}

For opposite detunings, $\Delta_1=-\Delta_2=\delta/2$, the
drift matrix of Eq.~\eqref{eq:drift_matrix} can be written as
$W=-i\mathcal H_{\mathrm{eff}}$, with the effective
non-Hermitian Hamiltonian
\begin{equation}
\mathcal H_{\mathrm{eff}}=
\begin{pmatrix}
\delta/2&-i\gamma_s/2\\
-i\gamma_s/2&-\delta/2
\end{pmatrix}
-\frac{i\Gamma}{2}\mathbb I_2 .
\label{eq:effective_hamiltonian}
\end{equation}
The coupling is purely imaginary, so $\mathcal H_{\mathrm{eff}}$
is anti-$\mathcal{PT}$ symmetric~\cite{Peng2016}. The
eigenvalues of $W$ are
\begin{equation}
\lambda_\pm=-\frac{\Gamma}{2}
\pm\frac12\sqrt{\gamma_s^2-\delta^2}.
\label{eq:reduced_eigenvalues}
\end{equation}
For $\delta\ne0$, the two eigenvalues and their
eigenvectors coalesce at $\gamma_s=|\delta|$, giving
\begin{equation}
\xi_{\mathrm{EP}}^{\mathrm{red}}
=\frac{|\delta|}{1+|\delta|}.
\label{eq:reduced_ep}
\end{equation}
Below this point, the modes have equal decay rates and
different oscillation frequencies. Above it, their
frequencies coincide in the rotating frame while their
decay rates split. The transition is driven entirely by the
shared dissipation; no direct coherent coupling between the
system resonators is needed. For the parameters of
Sec.~\ref{sec:parameters}, a mismatch $\delta=0.1$ corresponds
to $\delta/2\pi=50$~kHz, and the exceptional point is reached
at $g_c/2\pi\simeq1.1$~MHz.

\begin{figure*}[t]
\centering
\includegraphics[width=\textwidth]{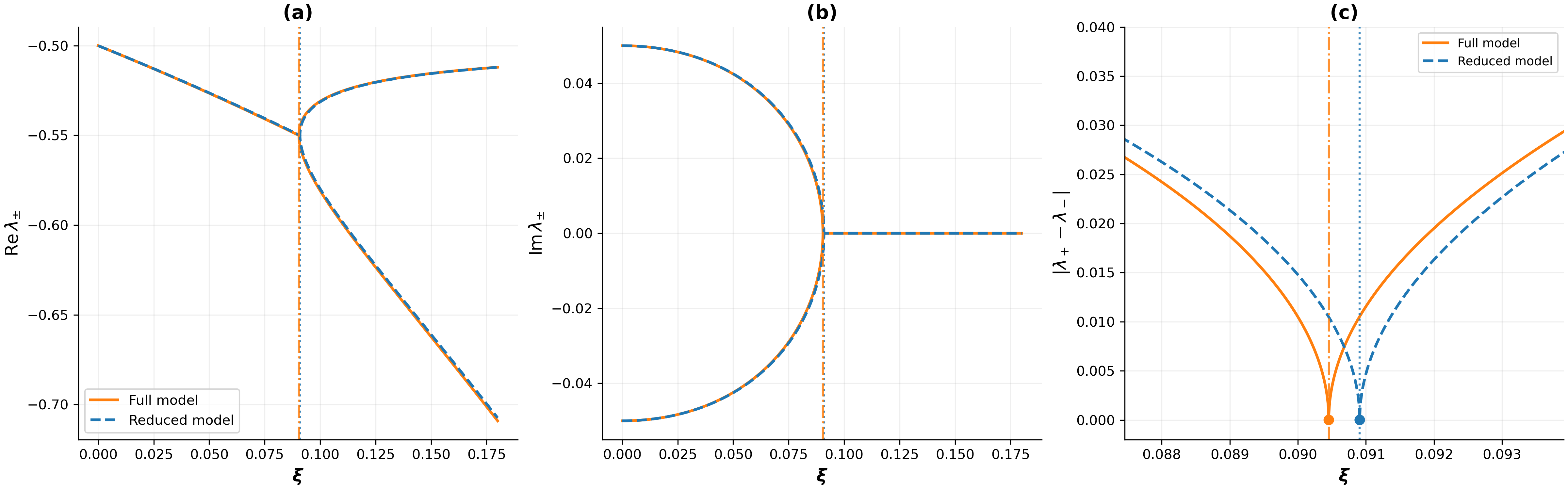}
\caption{
Exceptional-point structure of the full and reduced
models at fixed private decay. (a) Real and (b) imaginary
parts of the two slow eigenvalues as functions of $\xi$.
(c) Eigenvalue splitting $|\lambda_+-\lambda_-|$ near
the coalescence points. Solid orange and dashed blue
curves denote the full and reduced models, respectively.
The vertical orange dash-dotted and blue dotted lines
mark their respective exceptional points. Parameters
are $\kappa_b=200$ and $\delta=0.1$
($\Delta_1=-\Delta_2=0.05$). Eigenvalues are in units of
$\gamma_d$, and $g_c$ is varied according to
Eq.~\eqref{eq:xi_coupling}.
}
\label{fig:ep_spectrum}
\end{figure*}

To assess the elimination, we compare these eigenvalues
with the two slow branches of the full three-mode drift
matrix,
\begin{equation}
W_3=\begin{pmatrix}
-1/2-i\delta/2&0&-ig_c\\
0&-1/2+i\delta/2&-ig_c\\
-ig_c&-ig_c&-\kappa_b/2
\end{pmatrix},
\label{eq:full_drift}
\end{equation}
with $g_c$ given by Eq.~\eqref{eq:xi_coupling}, so that the
same value of $\xi$ specifies corresponding full and reduced
parameters. Figure~\ref{fig:ep_spectrum} shows close
agreement between the slow branches. The finite response
time of the auxiliary mode shifts the coalescence point
slightly, from $\xi_{\mathrm{EP}}^{\mathrm{red}}=0.0909$ to
$\xi_{\mathrm{EP}}^{\mathrm{full}}=0.0905$, a relative shift
of $0.5\%$. This shift scales as $1/\kappa_b$; halving
$\kappa_b$ to $100$ doubles it. The rapidly decaying auxiliary
branch is not shown.

Above the exceptional point, the mean amplitudes lock to a
common frequency. The slower eigenvector, belonging to
$\lambda_+$, satisfies
\begin{equation}
\frac{x_2}{x_1}=-e^{i\varphi},
\qquad
\sin\varphi=\frac{|\delta|}{\gamma_s},
\label{eq:locked_phase}
\end{equation}
where $x_j=\langle a_j\rangle$. The two resonators therefore
oscillate with equal amplitudes and a fixed phase difference
$|\arg(x_2/x_1)|=\pi-\varphi$, which approaches antiphase for
$\gamma_s\gg|\delta|$ and reduces to $\pi/2$ at the
exceptional point. Below the exceptional point, the two
eigenmodes have different frequencies and the relative phase
does not settle. For identical resonators, $\delta=0$, the
locked state is exactly antiphase; we use this limit below
because it admits closed-form expressions. For couplings of
opposite sign, $\xi<0$, the eigenvalues are unchanged and
$x_2/x_1=e^{i\varphi}$, so the locked state approaches in-phase
motion instead.

\subsection{Mean-amplitude synchronization}
\label{sec:synchronization_definition}

For resonant modes, $\Delta_1=\Delta_2=0$, we introduce the
collective modes $a_\pm=(a_1\pm a_2)/\sqrt2$. Their equations are
\begin{equation}
\dot a_\pm=-\frac{k_\pm}{2}a_\pm+F_\pm,
\label{eq:collective_dynamics}
\end{equation}
where $F_\pm=(F_1\pm F_2)/\sqrt2$ and
\begin{equation}
k_+=1+2\gamma_s,\qquad k_-=1.
\label{eq:collective_decay_rates}
\end{equation}
The symmetric mode couples to the shared reservoir,
whereas the shared noise and damping cancel from the
antisymmetric mode. The latter is dark with respect to
the shared channel but still undergoes private decay.
This mechanism is the common-bath route to synchronization
studied in Refs.~\cite{Giorgi2012,Manzano2013}; here it
arises from an explicit circuit element whose strength is
set by the SQUID flux bias.

Since the input fields have zero mean, the collective
amplitudes $x_\pm=(x_1\pm x_2)/\sqrt2$ obey
\begin{equation}
x_\pm(\tau)=x_\pm(0)e^{-k_\pm\tau/2}.
\label{eq:collective_mean_solution}
\end{equation}
For $\gamma_s>0$ and $x_-(0)\ne0$, the faster decay of
the symmetric component leads to the antiphase relation
$x_1\simeq-x_2$. For couplings of opposite sign the roles of
$a_\pm$ are exchanged, the symmetric mode is dark, and the
resonators synchronize in phase; the expressions below then hold
with $x_1+x_2$ and $x_1-x_2$ interchanged, so that $\cot(\theta/2)$
is replaced by $\tan(\theta/2)$. We quantify the remaining symmetric
component by
\begin{equation}
R(\tau)=\frac{|x_1(\tau)+x_2(\tau)|}{|x_1(\tau)-x_2(\tau)|}
=\frac{|\alpha_1+\alpha_2|}{|\alpha_1-\alpha_2|}
 e^{-\gamma_s\tau},
\label{eq:synchronization_ratio}
\end{equation}
where $\alpha_j=x_j(0)$. For a tolerance $0<\epsilon<1$,
the synchronization time $\tau_{\mathrm{sync}}$ is the
earliest time after which $R\leq\epsilon$ remains
satisfied. For nonzero initial symmetric and antisymmetric
components,
\begin{equation}
\tau_{\mathrm{sync}}=
\max\left\{0,\frac{1}{\gamma_s}
\ln\left(\frac{|\alpha_1+\alpha_2|}
{\epsilon|\alpha_1-\alpha_2|}\right)\right\}.
\label{eq:synchronization_time_general}
\end{equation}
For $\alpha_1=1$ and $\alpha_2=e^{i\theta}$,
\begin{equation}
\tau_{\mathrm{sync}}(\theta)=
\max\left\{0,\frac{1}{\gamma_s}
\ln\left(\frac{|\cot(\theta/2)|}{\epsilon}\right)\right\}.
\label{eq:synchronization_time_phase}
\end{equation}
An initially antisymmetric mean has
$\tau_{\mathrm{sync}}=0$, while an initially symmetric
mean has no antisymmetric component and never reaches
the antiphase condition. If both means vanish, the
criterion is undefined.

This criterion describes antiphase alignment of decaying
mean amplitudes rather than sustained oscillations, and
$\epsilon$ sets the allowed symmetric admixture. At fixed
drift parameters and initial means, $\tau_{\mathrm{sync}}$
is independent of the bath occupations and of the initial
squeezing.

\begin{figure*}[t]
\centering
\includegraphics[width=0.88\textwidth]{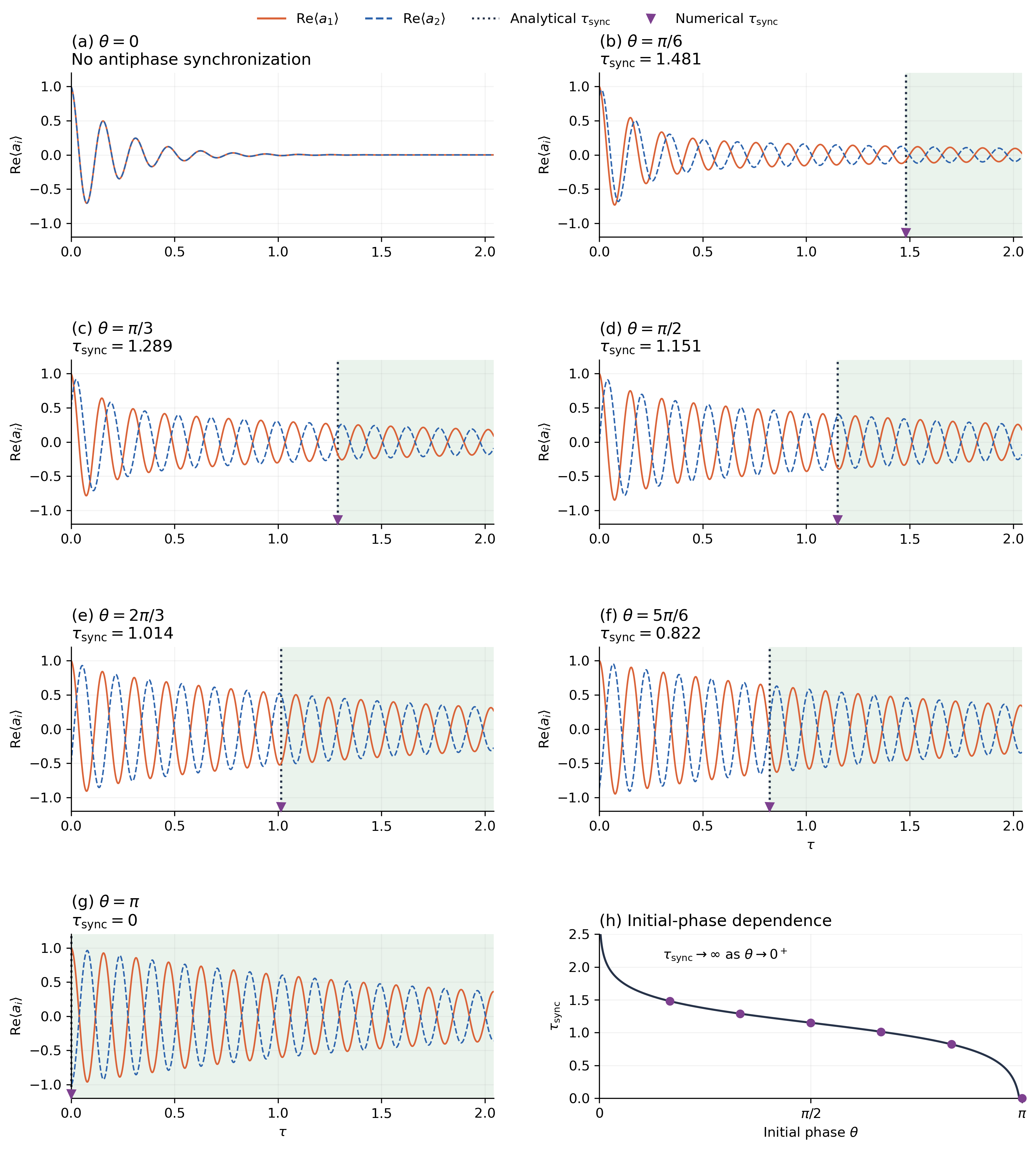}
\caption{
Initial-state dependence of antiphase synchronization
in the reduced model. (a)--(g) Evolution of
$\operatorname{Re}[x_1e^{-i\omega_r\tau}]$ (solid orange)
and $\operatorname{Re}[x_2e^{-i\omega_r\tau}]$ (dashed blue)
for $\alpha_1=1$ and $\alpha_2=e^{i\theta}$, with
$\theta=0,\pi/6,\pi/3,\pi/2,2\pi/3,5\pi/6,\pi$. The common
reference rotation $\omega_r=40$ is applied for visibility
only and does not affect $R$ or $\tau_{\mathrm{sync}}$.
Dotted vertical lines mark the analytical synchronization
times; triangles mark the numerically determined threshold
crossings. Shading indicates $R\leq\epsilon$.
(h) Analytical synchronization time versus the initial
phase (solid curve), with numerical values (markers).
Parameters are $\xi=0.8$ ($\gamma_s=4$, $\Gamma=5$),
$\Delta_1=\Delta_2=0$, and $\epsilon=0.01$. The reduced model
describes the device of Sec.~\ref{sec:parameters} with
$\kappa_b=200$; at fixed $\xi$, the curves do not depend on
$\kappa_b$.
}
\label{fig:synchronization}
\end{figure*}

Figure~\ref{fig:synchronization} illustrates the
initial-phase dependence at $\xi=0.8$. An initially
symmetric mean remains symmetric and decays rapidly,
whereas any nonzero antisymmetric component eventually
dominates. Increasing the initial phase difference toward
$\pi$ reduces the symmetric admixture and hence the
synchronization time. Numerical integration agrees with
Eq.~\eqref{eq:synchronization_time_phase}. These
trajectories are the same for coherent and displaced
squeezed initial states with the same $\alpha_1$ and
$\alpha_2$. For $\theta=5\pi/6$ and $\epsilon=0.01$,
$\tau_{\mathrm{sync}}\simeq0.82$, which corresponds to about
$0.26~\mu$s for the parameters of Sec.~\ref{sec:parameters}.

In the full three-mode model with $\kappa_b=200$,
$\tau_{\mathrm{sync}}$ is $3$--$4\%$ shorter for the initial
phases shown, because the exact decay rate of the symmetric
mode slightly exceeds $k_+$. This difference scales as
$1/\kappa_b$.

\subsection{Gaussian dynamics and entanglement}
\label{sec:gaussian_dynamics}

We introduce the quadratures
\begin{equation}
q_j=\frac{a_j+a_j^\dagger}{\sqrt2},
\qquad
p_j=\frac{a_j-a_j^\dagger}{i\sqrt2},
\label{eq:quadratures}
\end{equation}
with $[q_j,p_k]=i\delta_{jk}$, so the vacuum variance is
$1/2$. For $\boldsymbol u=(q_1,p_1,q_2,p_2)^T$, the
centered covariance matrix is
\begin{equation}
V_{jk}=\frac12\langle\delta u_j\delta u_k+
\delta u_k\delta u_j\rangle,
\qquad \delta u_j=u_j-\langle u_j\rangle.
\label{eq:covariance_definition}
\end{equation}
The system is initially prepared in a product of displaced
squeezed states,
\begin{equation}
|\psi(0)\rangle=
D_1(\alpha_1)S_1(s)|0\rangle\otimes
D_2(\alpha_2)S_2(s)|0\rangle,
\label{eq:initial_squeezed_product}
\end{equation}
where $D_j(\alpha_j)=\exp(\alpha_j a_j^\dagger-\alpha_j^*a_j)$
and $S_j(s)=\exp[s(a_j^2-a_j^{\dagger2})/2]$, with equal
real $s\geq0$ so that both modes are squeezed along the same
quadrature. The initial covariance is
\begin{equation}
V(0)=\frac12\operatorname{diag}
(e^{-2s},e^{2s},e^{-2s},e^{2s}),
\label{eq:initial_covariance}
\end{equation}
which does not depend on the displacements.

The linear dynamics with thermal Gaussian inputs preserve
Gaussianity~\cite{Weedbrook2012}, and the covariance evolves as
\begin{equation}
\dot V=AV+VA^T+D,
\label{eq:covariance_lyapunov}
\end{equation}
with
\begin{equation}
A=-\frac12\begin{pmatrix}
\Gamma&0&\gamma_s&0\\
0&\Gamma&0&\gamma_s\\
\gamma_s&0&\Gamma&0\\
0&\gamma_s&0&\Gamma
\end{pmatrix}.
\label{eq:quadrature_drift}
\end{equation}
For equal input occupations
$\bar n_1=\bar n_2=\bar n_b=n$, the diffusion matrix is
$D=-2vA$, with $v=n+1/2$. The stationary covariance is then
$v\mathbb I_4$, and
\begin{equation}
V(\tau)=v\mathbb I_4+
 e^{A\tau}[V(0)-v\mathbb I_4]e^{A^T\tau}.
\label{eq:covariance_exact_check}
\end{equation}

Entanglement is quantified by the logarithmic negativity
$E_N=\ln\|\rho_{12}^{T_2}\|_1$~\cite{VidalWerner2002}.
Partial transposition of mode 2 reverses $p_2$,
$\widetilde V=\Lambda V\Lambda$ with
$\Lambda=\operatorname{diag}(1,1,1,-1)$. Writing $V$ in
$2\times2$ blocks $B_1$, $B_2$, and $C$, and defining
$\widetilde\Sigma=\det B_1+\det B_2-2\det C$, the smallest
symplectic eigenvalue of $\widetilde V$ is
\begin{equation}
\widetilde\nu_-=
\sqrt{\frac{\widetilde\Sigma-
\sqrt{\widetilde\Sigma^2-4\det V}}{2}},
\label{eq:pt_symplectic_eigenvalue}
\end{equation}
and
\begin{equation}
E_N=\max\{0,-\ln(2\widetilde\nu_-)\}.
\label{eq:logarithmic_negativity}
\end{equation}
For two-mode Gaussian states, $\widetilde\nu_-<1/2$ is
necessary and sufficient for entanglement~\cite{Simon2000}.

The collective decay hierarchy that produces synchronization
also converts local squeezing into intermode entanglement. To
see this, we use the quadratures of the collective modes
$a_\pm$ introduced in Sec.~\ref{sec:synchronization_definition},
\begin{equation}
q_\pm=\frac{q_1\pm q_2}{\sqrt2},
\qquad
p_\pm=\frac{p_1\pm p_2}{\sqrt2}.
\label{eq:collective_quadratures}
\end{equation}
In the physical basis, each quadrature is damped at the rate
$\Gamma$ and coupled to the other resonator through the
cross-damping $\gamma_s$. In the collective basis, the
cross-damping adds to the symmetric rate and cancels from the
antisymmetric one, so $A$ becomes diagonal with entries
$-k_\pm/2$, where $k_\pm=\Gamma\pm\gamma_s$ are the rates of
Eq.~\eqref{eq:collective_decay_rates}. For equal occupations,
the collective noises are uncorrelated. Since both resonators
are squeezed identically, $V(0)$ has the same diagonal form in
the collective basis. The collective covariance therefore
remains diagonal, with variances
\begin{align}
v_{q,\pm}(\tau)&=
 v+\left(\frac{e^{-2s}}2-v\right)e^{-k_\pm\tau},\\
v_{p,\pm}(\tau)&=
 v+\left(\frac{e^{2s}}2-v\right)e^{-k_\pm\tau}.
\label{eq:collective_variances}
\end{align}
Partial transposition interchanges $p_+$ and $p_-$, so that
\begin{equation}
\widetilde\nu_-^2=
\min\{v_{q,+}v_{p,-},\,v_{q,-}v_{p,+}\}.
\label{eq:collective_entanglement_condition}
\end{equation}
Both products equal $1/4$ at $\tau=0$. Entanglement develops
when the anti-squeezed variance of the symmetric mode relaxes
quickly while the squeezed variance of the antisymmetric mode
degrades slowly, giving $v_{q,-}v_{p,+}<1/4$. Initial
squeezing supplies the nonclassical resource, and the shared
channel converts it into entanglement between the resonators.
For couplings of opposite sign, $a_+$ and $a_-$ exchange roles and
the relevant product becomes $v_{q,+}v_{p,-}$; since the two cases
are related by the local rotation $a_2\to-a_2$, $E_N$ is unchanged.

The disappearance of entanglement can be estimated in closed
form. For $k_+\gg1$, the symmetric mode reaches
$v_{p,+}\simeq v$ quickly, and the condition becomes
$v_{q,-}<1/(4v)$. This holds until
\begin{equation}
\tau_{\mathrm{d}}\simeq
\ln\frac{v-e^{-2s}/2}{v-1/(4v)}.
\label{eq:death_time}
\end{equation}
Since $\tau_{\mathrm{d}}$ depends only on the antisymmetric
mode, it is independent of $\xi$ once $k_+$ is large. For
$n=0$ ($v=1/2$), the denominator vanishes and the entanglement
decays only asymptotically.

The dynamics contain neither parametric pair creation nor a
squeezed reservoir. For coherent initial states ($s=0$),
passive linear evolution with thermal inputs keeps the
Glauber $P$ function positive and cannot generate entanglement.
For $s>0$ the entanglement is transient, since private damping
drives both collective modes to the same thermal covariance.
The role of squeezing is also visible in the centered
anomalous correlation $m_{12}=\langle\delta a_1\delta a_2\rangle$,
with $\delta a_j=a_j-\langle a_j\rangle$:
\begin{equation}
m_{12}(\tau)=\frac{\sinh(2s)}4
\left[e^{-\tau}-e^{-(1+2\gamma_s)\tau}\right].
\label{eq:anomalous_cross_correlation}
\end{equation}
A nonzero $m_{12}$ is not sufficient for entanglement,
because thermal fluctuations can render the state separable
while $m_{12}$ remains finite.

The onset of entanglement also follows from
Eq.~\eqref{eq:collective_entanglement_condition}. The product
$v_{q,-}v_{p,+}$ equals $1/4$ at $\tau=0$, and it falls below
$1/4$ immediately afterward when its initial slope is negative.
With $x=e^{2s}$, this condition reads
\begin{equation}
2v\,x^2-(1+k_+)\,x+2k_+v<0 .
\label{eq:onset_condition}
\end{equation}
The smaller root of the left-hand side gives the threshold
squeezing $s_{\mathrm{th}}$ above which entanglement appears.
For $\xi=0.8$, Eq.~\eqref{eq:onset_condition} gives
$s_{\mathrm{th}}=0.53$, $1.06$, and $2.13$~dB for $n=0.05$,
$0.1$, and $0.2$, in agreement with the numerical thresholds
to within $0.01$~dB. The quadratic has no real roots when
$v>(1+k_+)/(4\sqrt{k_+})$, that is, $n>0.33$ for $\xi=0.8$.
Numerically, we find no entanglement for $n=0.5$ at any
squeezing up to $17$~dB.

\begin{figure*}[t]
\centering
\includegraphics[width=\textwidth]{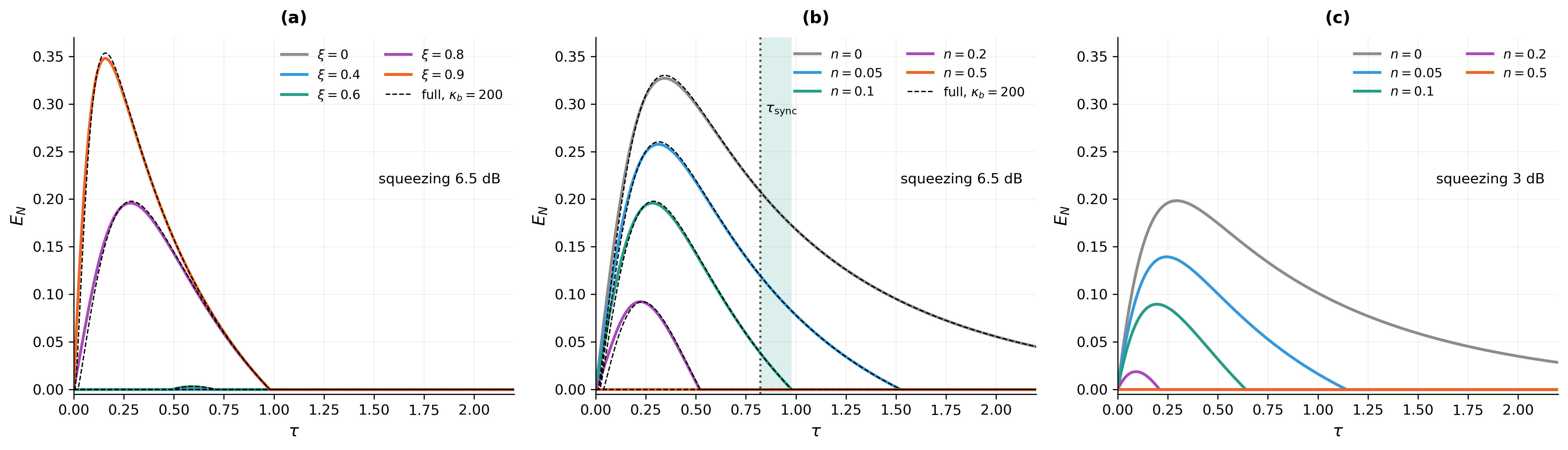}
\caption{
Transient entanglement generated by collective dissipation,
obtained from Eq.~\eqref{eq:covariance_lyapunov} (solid lines).
(a) Dependence on $\xi=0,0.4,0.6,0.8,0.9$ at fixed $n=0.1$,
with the private decay held fixed.
(b) Dependence on $n=0,0.05,0.1,0.2,0.5$ at fixed $\xi=0.8$.
(c) Same as (b) for $3$~dB of initial squeezing.
Panels (a) and (b) use $6.5$~dB ($s=0.75$) and panel (c)
$3$~dB ($s=0.35$) of squeezing, with identical squeezing
orientations and $\Delta_1=\Delta_2=0$. Dashed black lines in
(a) and (b) show the full three-mode model with
$\kappa_b=200$. In (b), the dotted line marks
$\tau_{\mathrm{sync}}$ for $\alpha_1=1$,
$\alpha_2=e^{i5\pi/6}$, and $\epsilon=0.01$, and the shaded
band marks the coexistence of $R\leq\epsilon$ and $E_N>0$ for
$n=0.1$. Displacements affect the synchronization marker but
not $E_N$.
}
\label{fig:entanglement}
\end{figure*}

Figure~\ref{fig:entanglement} shows $E_N$ for initial
squeezing of $6.5$~dB ($s=0.75$) and $3$~dB ($s=0.35$). At
fixed $n=0.1$ [Fig.~\ref{fig:entanglement}(a)], increasing
$\xi$ separates the collective decay rates and enhances the
peak entanglement: $\xi=0$ and $0.4$ give none, $\xi=0.6$
gives a weak transient, and $\xi=0.8$ and $0.9$ give
pronounced peaks. The $\xi=0.8$ and $0.9$ curves vanish at
the same time, $\tau\simeq0.98$, which coincides with
Eq.~\eqref{eq:death_time} because both $k_+=9$ and $k_+=19$
are large compared with $k_-=1$.

At fixed $\xi=0.8$ [Fig.~\ref{fig:entanglement}(b)],
increasing $n$ raises the diffusion without changing the
drift. Both the peak entanglement and its lifetime decrease,
and no entanglement is obtained for $n=0.5$.
Equation~\eqref{eq:death_time} gives
$\tau_{\mathrm{d}}\simeq1.53$, $0.98$, and $0.54$ for
$n=0.05$, $0.1$, and $0.2$, close to the numerical values
$1.52$, $0.98$, and $0.52$. For the parameters of
Sec.~\ref{sec:parameters}, $\tau_{\mathrm{d}}\simeq0.98$
corresponds to about $0.31~\mu$s.

At $5$~GHz, $n=0.1$ corresponds to an effective temperature
of about $100$~mK. Superconducting resonators in dilution
refrigerators typically thermalize near $40$--$60$~mK, where
$n\lesssim0.02$. The value $n=0.1$ is therefore conservative,
and Fig.~\ref{fig:entanglement}(b) shows that the
entanglement survives at occupations an order of magnitude
above the expected value; Eq.~\eqref{eq:onset_condition} sets
the limit at $n\simeq0.33$ for $\xi=0.8$. For $n=0.02$, Eq.~\eqref{eq:death_time} gives
$\tau_{\mathrm{d}}\simeq2.3$, or about $0.75~\mu$s.

With $3$~dB of initial squeezing
[Fig.~\ref{fig:entanglement}(c)], the entanglement is weaker
and shorter-lived, with a peak $E_N\simeq0.09$ that vanishes
at $\tau\simeq0.64$ for $n=0.1$. Since $3$~dB lies above
the threshold for $n\leq0.2$, entanglement persists for all
occupations except $n=0.5$, although it becomes weak for
$n=0.2$.

The full three-mode covariance dynamics with $\kappa_b=200$,
including the auxiliary mode $b$, reproduce these results
[dashed lines in Figs.~\ref{fig:entanglement}(a)
and~\ref{fig:entanglement}(b)]. For $\xi=0.8$ and $0.9$ and
all occupations shown, the peak $E_N$ differs by less than
$2\%$ and the time at which entanglement vanishes agrees to
within $0.01$. The only visible difference is a short initial
delay, during which the auxiliary mode builds up its
correlations with the system modes.

The synchronization time is unaffected by $n$, because the
first moments do not depend on the occupations. For
$\theta=5\pi/6$ and $\epsilon=0.01$,
$\tau_{\mathrm{sync}}\simeq0.82$, so at $n=0.1$ antiphase
alignment and entanglement coexist over
$0.82\lesssim\tau\lesssim0.98$ [shaded in
Fig.~\ref{fig:entanglement}(b)]. The width of this window
depends on $\epsilon$ and $\theta$. The two effects share
their origin in the collective decay hierarchy but are
distinct: synchronization concerns the first moments, whereas
$E_N$ concerns the centered fluctuations. Coherent initial
states show the same synchronization while remaining
separable.

\section{Experimental considerations}
\label{sec:experimental}

\subsection{Preparation of the squeezed initial state}
\label{sec:state_preparation}

The entanglement results of Sec.~\ref{sec:gaussian_dynamics}
require each system resonator to be squeezed before $\tau=0$.
This can be done with a flux-pumped SQUID in each system
resonator, as in flux-driven Josephson parametric
amplifiers~\cite{Yamamoto2008}. Modulating its flux at
$2\omega_j$ generates the squeezing Hamiltonian
$i(\epsilon_j/2)(a_j^{\dagger2}-a_j^2)$, and a pump pulse of
duration $T_p$ produces $s=\epsilon_jT_p$ in the absence of loss.
The pump is switched off before $\tau=0$. Afterward the flux
through this SQUID is static, so it only renormalizes $\omega_j$,
and the dynamics remain those of the time-independent model of
Sec.~\ref{sec:model}.

In the steady state, intracavity squeezing generated by a
parametric drive on a damped mode is limited to $3$~dB. A short
pump pulse with $\epsilon_j\gg\gamma_d$ avoids this limit,
because the private loss acts only weakly during the pulse.
Intracavity squeezing beyond $3$~dB has also been demonstrated
by reservoir engineering~\cite{Dassonneville2021}. For the
parameters of Sec.~\ref{sec:parameters}, a pump rate
$\epsilon_j/2\pi=5$~MHz reaches $6.5$~dB in about $26$~ns,
including the private loss during the pulse. This is an order
of magnitude shorter than $1/\gamma_d\simeq0.32~\mu$s, and the
resulting state has a purity of about $0.95$. A pump rate of
$3$~MHz requires about $45$~ns.

During the preparation, the couplers should be biased at a
weakly coupled point, so that the shared channel does not act
on the resonators while they are being squeezed. Stepping the
flux biases to the operating point then starts the dynamics at
$\tau=0$. The step switches between two static configurations,
and the auxiliary mode relaxes to its thermal state within
$2/\kappa_b\simeq3$~ns. This is the initial condition used in
the full three-mode calculations of Sec.~\ref{sec:results}.

\subsection{Readout}
\label{sec:readout}

Both quantities studied in Sec.~\ref{sec:results} are obtained
from the fields leaving the private ports,
Eq.~\eqref{eq:input_output}. Averaged heterodyne records give
$\langle a_{j,\mathrm{out}}\rangle=\sqrt{\gamma_{d,j}}\,x_j$,
and hence $x_1\pm x_2$ and $R(\tau)$. For equal private decay
rates, $R$ is independent of $\gamma_d$ once the gain and phase
of the two detection chains are calibrated against each other.

At the synchronization threshold the antiphase signal is still
large. The antisymmetric amplitude has decayed only by the
factor $e^{-k_-\tau_{\mathrm{sync}}/2}$, which equals $0.66$ for
$\theta=5\pi/6$. The demanding part of the measurement is
resolving the symmetric admixture at the level
$R\simeq\epsilon=0.01$, which requires averaging over many
repetitions. Since the resonators return to equilibrium within
a few microseconds, $10^5$ repetitions take less than one second.

The covariance matrix, and hence $E_N$, follows from the second
moments of the amplified output records. The added noise of the
amplification chain is subtracted using reference measurements
with the resonators in their thermal
state~\cite{Eichler2011b,Eichler2012}.
Near-quantum-limited amplifiers~\cite{Macklin2015} reduce the
number of repetitions required. Reconstructing the
\emph{intracavity} covariance $V(\tau)$ requires temporal mode
matching: the output records are integrated with filter
functions matched to the resonator modes, and the measured
moments are related to the intracavity moments through
Eq.~\eqref{eq:input_output}~\cite{Eichler2012}. Because $V(\tau)$
evolves on the timescale $1/k_+$, the filter windows must be
short compared with it, or the measured output correlations can
be compared directly with the model.

The dynamics last about $0.3~\mu$s, and their fastest timescale
is the decay of the symmetric mode, at rate
$k_+\gamma_d/2\pi\simeq4.5$~MHz for $\xi=0.8$. A detection
bandwidth of several megahertz is therefore sufficient. This is
well within the bandwidth of current readout chains, including
traveling-wave parametric amplifiers~\cite{Macklin2015}. The
strongly damped auxiliary port does not need to be measured.

\subsection{Imperfections}
\label{sec:imperfections}

\textit{Coupling asymmetry.} For equal private decay and
$g_1\neq g_2$, the combination $g_2a_1-g_1a_2$ remains an exact
eigenmode of the reduced drift matrix, decaying at the private
rate. The long-time state is then still antiphase, but with
amplitude ratio $|x_2/x_1|=g_1/g_2$. Unequal private decay
rates, $\gamma_{d,1}\neq\gamma_{d,2}$, mix this mode with the
bright mode; the admixture is small when
$|\gamma_{d,1}-\gamma_{d,2}|\ll\gamma_{s,1}+\gamma_{s,2}$.

\textit{Residual detuning.} A residual frequency mismatch
$\delta$ above the exceptional point shifts the locked relative
phase from $\pi$ to $\pi-\arcsin(|\delta|/\gamma_s)$, according
to Eq.~\eqref{eq:locked_phase}. For $\xi=0.8$ ($\gamma_s=4$) and
$\delta=0.1$, the shift is $0.025$~rad.

\textit{Flux noise.} Away from zero flux, the SQUID inductance
depends on $\Phi_{j,\mathrm{dc}}$ to first order, so flux noise
modulates both $g_j$ and the resonator frequencies. Since the
relevant dynamics last a few hundred nanoseconds, slow flux
drift mainly produces shot-to-shot variation of $\xi$ and of the
detunings. This variation can be reduced by operating close to
zero flux, where the inductance is first-order insensitive.
\section{Conclusion}
\label{sec:conclusion}

We have proposed a superconducting circuit in which two
resonators share a dissipation channel through a strongly
damped auxiliary resonator, with the couplings set by static
flux biases of two DC SQUIDs. Adiabatic elimination of the
auxiliary mode yields cross-damping and correlated noise that
originate from the same collective dissipator. The magnitude of
the shared-channel parameter $\xi$ is tuned continuously within
$0\leq|\xi|<1$ by the coupler biases, while its sign is fixed by
the relative sign of the two couplings. The circuit thus realizes
the correlated dissipation of Ref.~\cite{BittnerTyagi2025}
without a direct coherent coupling between the resonators.

The shared channel produces three effects. For detuned
resonators, the eigenvalues coalesce at an exceptional point
$\xi_{\mathrm{EP}}=|\delta|/(1+|\delta|)$, above which the mean
amplitudes lock to a common frequency with a fixed relative
phase. For identical resonators, the collective mode that couples
to the auxiliary resonator decays faster than the dark mode, which
drives the mean amplitudes into antiphase alignment for couplings
of equal sign and into in-phase alignment for couplings of opposite
sign. The synchronization time is set by the collective decay
splitting and the initial mean amplitudes. When the
resonators are initially squeezed, the same decay hierarchy
converts local squeezing into transient entanglement. We
obtained closed-form expressions for the squeezing threshold
above which entanglement appears and for the time at which it
disappears. Synchronization and entanglement share this origin
but remain distinct: the former concerns first moments and
also occurs for coherent states, whereas the latter requires
initial squeezing and is limited by thermal noise. The full
three-mode dynamics confirm the reduced description to within a
few percent for $\kappa_b=200$. For a private decay rate
$\gamma_d/2\pi=0.5$~MHz, these effects occur within about
$0.3~\mu$s, persist at thermal occupations an order of magnitude
higher than those of typical devices, and are accessible through
the output fields of the private ports.

Beyond realizing the Bittner--Tyagi model, the circuit is useful
in its own right because the noise correlation is set by the
coupler fluxes and, for identical modes, can be calibrated from
the two collective ring-down rates, $|\xi|=(k_+-k_-)/(k_++k_-)$.
It can therefore serve as a controllable source of correlated
noise, a test bed for the correlated errors that limit error
correction in superconducting processors~\cite{Wilen2021,McEwen2022}.
The dark mode, which decays only through the private losses, is a
bosonic analogue of a decoherence-free subspace~\cite{Lidar1998}
whose protection can be switched with the flux. Since the
entanglement leaks out through the two independent lines, the
device can also serve as a source of entangled microwave fields
that requires no two-mode pump~\cite{Flurin2012}.
\section*{Data availability}
The code and data that support the findings of this article will be made openly available upon publication.

\begin{acknowledgments}
This work was supported by the National Science Foundation under Grant No. CHE-2404788 and the Robert A. Welch Foundation (Grant No. E-1337).
\end{acknowledgments}

\section*{Author contributions}

Areeda Ayoub conceived the circuit-based approach to
engineering thermal correlations, developed the model,
performed the analytical and numerical calculations,
prepared the figures, and wrote the initial manuscript.
Eric Bittner provided scientific guidance, helped
refine the research direction, and supervised the project.
Both authors contributed to interpreting the results
and reviewing and editing the manuscript.

\section*{Declaration of AI assistance}

During this work, the authors used ChatGPT (OpenAI) and Claude
(Anthropic) to assist with discussions of the theoretical model,
checking analytical derivations, developing and reviewing numerical
and plotting code, verifying references, and drafting and editing
the manuscript. All AI-assisted material was critically reviewed and
verified by the authors, who take full responsibility for the
accuracy and integrity of the reported results and the final
manuscript.

\section*{Conflicts of Interest}
The authors declare no competing financial or non-financial interests.

\bibliographystyle{apsrev4-1}
\bibliography{bibliography-paper}

\end{document}